\magnification1200

\rightline{KCL-MTH-10-08}

\vskip .5cm
\centerline
{\bf  $E_{11}$, generalised space-time and IIA string theory}
\vskip 1cm

\centerline{Peter West}
\centerline{Department of Mathematics}
\centerline{King's College, London WC2R 2LS, UK}

\vskip .2cm
\noindent 
As advocated in hep-th/0307098 we construct the non-linear realisation of
the semi-direct product of $E_{11}$ and its first fundamental
representation at lowest level from the IIA viewpoint. We find a theory
that is $SO(10,10)\otimes GL(1)$ invariant and contains the fields of
gravity, a two form and a dilaton but which depend on coordinates
which belong to the vector representation of SO(10,10). The resulting
Lagrangian agrees that of recent work on the  so called doubled field
theory. However, the construction given in this paper is straightforward
and  systematic. It also reveals the relevant underlying symmetries and
opens the way to include the Ramond-Ramond, and higher level, fields
together with  additional coordinates of the generalised space-time. 

\vskip .5cm

\vfill
\eject

When the $E_{11}$ symmetry was first conjectured space-time
was encoded  into the non-linear realisation by introducing the
space-time  translation generators in an ad hoc manner [1,2]. It was
subsequently proposed [3] that one should introduce generators
transforming in  the fundamental 
 representation $l_1$  of $E_{11}$; more precisely one should  
take the non-linear realisation of the semi-direct product of $E_{11}$
and a set of generators in the
fundamental  representation associated with node one ( see figure 1) ,
i.e.
$E_{11}\otimes_s l_1$ [3]. At lowest levels the $l_1$  multiplet in
eleven dimensions  begins with the space-time translation generators
$P_a$,  then a two form $Z^{a_1a_2}$,  a five form generator 
$Z^{a_1\dots a_5}$ and a generator $Z^{a_1\dots a_7,b}$ together with an
infinite number of other generators.  In this approach the
fields in eleven dimensions would depend on all the coordinates introduced
in this non-linear realisation   that is
$x^a, x_{a_1a_2}, x_{a_1\ldots a_5}, x_{a_1\dots a_7,b},\ldots$ [3]. 
\par
There is convincing evidence that all the brane charges are contained in
the $l_1$ representation [3,4,5,6,7] and the introduction of the
different coordinates corresponds to measuring using the different
brane probes.  The non-linear realisation based on $E_{11}\otimes_s l_1$
is a theory that possess many new unfamiliar coordinates, indeed at all
levels an infinite number of coordinates and it is not clear how to
recover a field theory which has only the 
dependence on the usual space-time. Nonetheless the non-linear
realisation of
$E_{11}\otimes_s l_1$  has been used to construct the gauged
supergravities with maximal supersymmetry, at least the field strengths
of the four dimensional theories, although the techniques used are apply
to all dimensions [8]. In this case not all the coordinates of the $l_1$
representation were used but instead of taking just the usual space-time
coordinates $x^a$ one took a space-time to be a slice which lies in $l_1$
and $E_{11}$ and  contains many new coordinates. 
\par
A new generalised geometry  was proposed in references [8,9] and then
subsequent to reference [3]  used in a large number of papers to
reformulate parts of supergravity theories. However, reference [3] 
automatically introduces a generalise tangent space with a generalised
vierbein which is easily computed and whose role in the theory is very
strongly constrained by the symmetries of the non-linear realisation.  It
is clear that many features of the work using  generalised geometry  are
automatically encoded in the non-linear realisation of
$E_{11}\otimes_s l_1$. 
\par
In this paper we will compute the non-linear realisation of
$E_{11}\otimes_s l_1$ at the lowest level appropriate for the IIA theory.
This has a 
$SO(10,10)\otimes GL(1)$ symmetry and the same fields as the
NS-NS sector of the superstring, however, these fields depend on the
coordinates $x^a,y_a$ that belong to the vector representation of
SO(10,10). Following the ideas put forward in [11,1,2,12] we consider the
fields to only depend on $x^a$ and  demand that the theory is invariant
under diffeomorphisms. This fixes the coefficients in the Lagrangian and
we arrive at the Lagrangian for the Neveu-Schwarz-Neveu-Schwarz sector of
IIA supergravity. 
A key role is played by the $GL(1)$ part of the symmetry without which a
group theoretic derivation of the result would not have been possible. 
The coordinates $x^a,y_a$ were first introduced in the context of string
theory in reference [13,14]. In a recent paper [15] the the non-linear
realisation of
$E_{11}\otimes_s l_1$ at lowest level appropriate for the IIA theory was
used to deduce a formulation [13] of the bosonic sector of the
superstring.  
\par
To find a ten dimensional theory from the $E_{11}$  non-linear
realisation one must select an $A_9$, or Sl(10), subalgebra, the so called
gravity line,  as this subalgebra leads  to ten dimensional
gravity. Looking at the Dynkin diagram of $E_{11}$, see figure 1, one
sees that there are only two possibilities; the nodes from one to nine
inclusive and the nodes one to eight inclusive and node eleven. The
latter leads to the IIB theory [16] and the former the IIA theory [2]
which is the subject of this paper. There are two nodes in the IIA theory
which are  not associated with gravity, that is nodes
ten and eleven. Deleting node ten (see fig 1) leaves the algebra
$D_{10}$, in particular its real form SO(10,10),  and it is useful to
decompose the
$E_{11}$ adjoint representation,  which is the one that occurs in the
non-linear realisation,  into
representations of SO(10,10). However, to recognise objects that are more 
familiar it is helpful  to further decompose these representations 
into those of SL(10) which corresponds to deleting  node
eleven in addition. 
\par
The representations that  occur  in a  decomposition of the adjoint
representation of $E_{11}$ to a subalgebra associated with a deleted node
can be classified  in terms of increasing levels, whose precise
definition is given in appendix A.  The decomposition into
representations of SO(10,10), associated with the deletion of node ten,
at level zero  is just the adjoint representation of SO(10,10) together
with one other generator which is in the Cartan subalgebra of $E_{11}$. 
When written in terms of representations of SL(10) the generators of the
adjoint representation of  SO(10,10) are
$K^a{}_b, \ R^{ab},\ \tilde R_{ab}, \ a,b=1,2\ldots ,10$, where the 
$K^a{}_b$ are the generators of the adjoint representation of the SL(10)
of the chosen gravity line. There other generator  we will denote as
$\tilde R$.  Their algebra is derived from
$E_{11}$ in appendix A and is given by 
$$
[K^a{}_b,K^c{}_d]=\delta _b^c K^a{}_d - \delta _d^a K^c{}_b,  \ \ 
 [K^a{}_b, R^{c d}]=\delta_b^c R^{ad}-\delta_b^d R^{ac},
[K^a{}_b, \tilde R_{c d}]=-\delta_c^a \tilde R_{bd}+\delta_d^a \tilde
R_{bc},
$$
$$
 [R^{ab}, \tilde R_{c d}]=4\delta_{[c}^{[a}
K^{b]}{}_{d]}+{2\over 3} \tilde  R,\ \ \ [R^{ab}, R^{c d}]=0=[\tilde
R_{ab},
\tilde R_{c d}],\ [\tilde R,  R^{c d}]=0,\ [\tilde R,  \tilde R_{c d}]=0
\eqno(1) $$
where $\tilde R= - \sum_{a=1}^{11}K^a{}_a +3K^{11}{}_{11}$. We
will  show in appendix A that this algebra is  that of
$SO(10,10)\otimes GL(1)$ where the last factor has the generator  $\tilde
R$. 
\par
In the non-linear realisation of $E_{11}$ the generators of the
Borel subgroup  lead to fields and so at level zero we find 
  the fields
$h_a{}^b$,
$A_{a_1a_2}$ and $a$ that is the fields of the NS-NS sector of the
IIA string. At the next  level, i.e. level one, one finds generators, and
so fields,  which are all anti-symmetric tensors of SL(10) which have  
odd   ranks, i.e. those of  one to nine and these belong to the Majorana
Weyl spinor representation of SO(10,10).  These include the fields of the
Ramond-Ramond sector  and their duals as well as a rank nine form field
that   gives rise to the massive IIA theory. At higher levels one finds
infinite number of  fields many of which have a complicated index
structure.  In fact the six form field which is the dual of the  two
form field at level zero  occurs at level two. 
\par
As explained in reference [3] to introduce a generalised space-time into
the non-linear realisation we must also consider the fundamental
representation of
$E_{11}$ associated with node one, denoted by $l_1$, see appendix A. In
particular we will consider the non-linear realisation of the 
semi-direct product of $E_{11}$ and a set of generators that transform
under $E_{11}$ transformations, i.e
$E_{11}$ commutators, like the
$l_1$ representation; we denote this semi direct product as 
$E_{11}\otimes_s l_1$. The coordinates of the generalised space-time
arise as the parameters of the $l_1$ generators in the general group
element.  At level zero
with respect to node ten the generators of the $l_1$ representation are
$l_1^0=\{ P_a, Q^a, \ a=1,2\ldots ,10\} $ which belong to the vector
representation of 
$SO(10,10)$. Their commutation relations with those of  $SO(10,1))\otimes
GL(1)$ are derived in appendix A from the $E_{11}\otimes_s l_1$ algebra
and are given by  
$$
[K^c{}_b, P_a]=-\delta_a^c P_b+{1\over 2} \delta ^c_bP_a,\ \ 
 [  R^{ab}, P_c]=-{1\over 2}
(\delta^a_c Q^b-\delta^b_c Q^a)=- \delta^{[a}_c Q^{b]},\ \  [\tilde  R_{ab}, P_c]=0,
$$
$$
[K^a{}_b, Q^c]=\delta_b^c Q^a+{1\over 2} \delta ^c_b Q^a\ \  [ \tilde R_{ab},  Q^c]=4 (\delta_a^c P_b-\delta_b^c P_a)= 2 \delta_{[a}^c P_{b]},\ \  [R^{ab},Q^c]=0,
$$
$$
[\tilde R, P_a]= -3P_a,\ [\tilde R, Q^a]= -3Q^a
\eqno(2)$$
\par
At  level one the generators of $l_1$, and so the 
associated coordinates, are all tensors of SL(10) of  even
rank and they belong to the  Majorana Weyl spinor representation of
SO(10,10). It is of opposite chirality to those of the level one
fields mentioned above. 
\par
We can now construct the non-linear realisation of  
$E_{11}\otimes _s l_1$ with local sub-algebra $K(E_{11})$ at level zero
where $K(G)$ denotes  the Cartan involution invariant subalgebra of the
algebra $G$. This is just the non-linear realisation of $(E_{11}\otimes
_s l_1)^{(0)}= (SO(10,10)\otimes GL(1))\otimes_s l_1^0$, whose algebra is
given in equations (1) and (2) above. The  local subalgebra of the
non-linear realisation is just the Cartan involution invariant subalgebra
of SO(10,10) which is 
$SO(10)\otimes SO(10)$. The  non-linear realisation is built from a
general element of $(E_{11}\otimes
_s l_1)^{(0)}$ which  is taken to transform as
$g\to g_0 g h$ where  the rigid transformation $g_0\in (E_{11}\otimes_s
l_1)^{(0)}$ and the local transformation 
$h\in SO(10)\otimes SO(10)$. Using the latter we can bring  the general
group element of $(E_{11}\otimes_s l_1)^{(0)}$ to be of the form 
$$
g=g_lg_E, \quad {\rm where } \quad g_l= e^{x^a P_a+ y_a Q^a} 
\quad {\rm and } \quad g_E=e^{h\cdot K} e^{{1\over 2}A\cdot R} e^{aR}
\eqno(3) $$
where $R={1\over 12} (9K^{11}{}_{11} -\sum_{a=1}^{11}K^a{}_a )$  and the
fields $h_a{}^b, A_{a_1a_2}$ and $a$ depend on $x^a$ and $y_a$. We note
that we are using a different scalar generator $R$ rather than the
$\tilde R$ that appeared in equation (1) and it has the commutators 
$$
[R, P_a]= 0,\ [R, Q^a]= {1\over 2}Q^a,\ [R,K^a{}_b ]=0,\ [R,
R^{ab}]={1\over 2}R^{ab} , 
\ [R, R_{ab}]=-{1\over 2}R_{ab} 
\eqno(4)$$
\par
The Cartan form is given by 
$$
{\cal V}= g^{-1} d g= g_E^{-1}(dx^aP_a+ dy_aQ^a) g_E+ g_E^{-1}d g_E
\eqno(5) $$ 
The first term is given by 
$$
g_E^{-1}(dx^aP_a+ dy_aQ^a) g_E\equiv 
(dx^a, dy_a) {\cal E}  \left( \matrix {P_b\cr Q^b}\right)= dz^T {\cal E} L
\eqno(6)$$
where the $2D$ by $2D$ matrix ${\cal E}$ is given by 
$$
{\cal E}= (det e)^{-{1\over 2}}\left(\matrix {e&{1\over 2} eAe^{-{1\over 2}a}\cr
0&e^{-1T}e^{-{1\over 2}a}\cr}\right)
\eqno(7)$$
where $e=e_\mu{}^a= (e^h)_\mu{}^a$ and $A$ is  the matrix $A_{a_1a_2}$.
We can think of ${\cal E}$ as a  generalised vielbein. In the last
line we have use the definitions  $dz^{T}=(dx^a, dy_a)$ and $L= \left(
\matrix {P_b\cr Q^b}\right)$. We note that ${\cal E}$ does not  have
determinant one, but this  is consistent with the presence of the
additional GL(1) generator in the algebra whose non-linear realisation we
are constructing.
\par
The Cartan form is inert under the above rigid transformations
but transforms  under the local  transformation
$h$ as
${\cal V}\to h^{-1} {\cal V}h+h^{-1}d h$. However, carrying out a rigid
transformation $g\to g_0 g$ on the general group element of equation (3)
we find that  $g_l \to g_0 g_l g^{-1}_0$ and $g_E\to g_0g_E$ as 
 the $l_1$ generators form a representation of $E_{11}$. As a result the 
coordinates form a representation of
$E_{11}$,  and in particular at level  zero $x^a$ and $y_a$ form a
representation of
$E_{11}^{(0)}\equiv SO(10,10)\otimes GL(1)$ and the action of rigid
$E_{11}^{(0)}$  transformations on the coordinates is given by 
$$
{x^a P_a+ y_a Q^a} \to {x^{a\prime } P_a+ y_a^\prime  Q^a} = g_0 (x^a P_a+
y_a Q^a) g_0^{-1}
\eqno(8)$$
As expected  they transform according to the vector representation of
SO(10,10). 
\par
We can  define the action of the vector  representation of
$k\in E_{11}^{(0)}= SO(10,10)\otimes GL(1)$ using the generators of
$l_1^{(0)}$  by
$U(k)( L_N )\equiv k^{-1} L_N k= D(k)_N{}^M L_M$ where 
$L_N= \left( \matrix {P_b\cr Q^b}\right)$ and $D(k)_N{}^M $  is the
matrix representative. As a result of equation (8), we find that
$dz^T
\to dz^{T\prime}= dz ^T D(g_0^{-1})$ or putting in the indices $dz^{T N}
\to dz^{T N\prime }= dz^{T M}D(g_0^{-1})_M{}^N$ where as before $dz^{T
N}= (dx^a, dy_a)$.  The derivative
$\partial_N= {\partial\over \partial z^N}$ in the generalised space-time
transforms as $\partial_N^\prime= D(g_0)_N{}^M\partial_M$. 
Examining equation (5), we note that
${\cal E}$ is equal to the matrix $D(g_E)_N{}^M$. Almost
identical considerations hold  at all levels in the $E_{11}\otimes_s
l_1$ non-linear realisation. 
\par
As the Cartan form is inert under rigid transformations,  their action on
the coordinates must be  compensated by that on ${\cal E}$ which we give
the indices ${\cal E}_N{}^A$ and so ${\cal E}_N{}^{A\prime} =
D(g_0)_N{}^M {\cal E}_M{}^{A}$. The generalised vielbein ${\cal
E}_N{}^{A\prime} $ transforms on its upper index by a local 
$SO(10)\otimes SO(10)$ transformation and so we can  think of the upper
index as a tangent index and the lower index as a world index. 
\par
Rather
than use the Cartan forms we will construct the action out of $M\equiv
g_EI_c(g_E^{-1})$  where $I_c$ is the Cartan involution. It is easy to see
that $M$ is inert under local transformations but transforms as $M\to
M^\prime = g_0 M I_c(g_0^{-1})$ under rigid transformations. Fortunately
at level zero, that is for the group $E_{11}^{(0)}$,  $I_c(g^{-1})=g^T$. 
It follows from the above  discussion just above, and equation (7)  that
in the  vector representation $M$ takes the form 
$$
D(M)={\cal E}{\cal E}^T= (det e)^{-1}
\left(\matrix { ee^T-{1\over 4} eAAe^T e^{-a}& {1\over 2}
eAe^{-1}e^{-a}\cr -{1\over 2} e^{-1T}Ae^{T}e^{-a}&
e^{-1T}e^{-1}e^{-a}}\right)
\eqno(9)$$ 
Writing out the indices explicitly $ D(M)_{NM}=({\cal E}{\cal E}^T)_{NM}$
it is clear that $D(M)$  is inert under local $SO(10)\otimes SO(10)$
transformations as the tangent index is summed  but transforms as a
SO(10,10)  vector on both of its lower world indices. In what follows we
write
$D(M)$ just as $M$ for simplicity. 
\par
We can construct an invariant Lagrangian out of $M$ and $\partial_N$.
These   are inert under local transformation and so the most general
such object bilinear in generalised space-time derivatives is given by 
$$
L=c_1 \partial_S M_{PQ} \partial_T (M^{-1})^{PQ}(M^{-1})^{ST}
+c_2 \partial_S (M^{-1})^{PQ}\partial_P M_{TQ} (M^{-1})^{ST}
$$
$$
+c_3 \partial_S (M^{-1})^{SR}\partial_P (M^{-1})^{PQ} M_{RQ}
+{c_4\over 400} \partial _S det M \partial _T (det M )^{-1}
(M^{-1})^{ST}
$$
$$
+{c_5\over 10}  det M \partial _T  (det M )^{-1} \partial _S
(M^{-1})^{ST} 
\eqno(10)$$
where $c_1,\ldots , c_5$ are constants. 
\par
It is instructive to evaluate  this Lagrangian in terms of the fields
and this is easiest achieved by going to string frame. That is carrying
out the field redefinitions 
$e\to e^{-{1\over 4}a} e_s, A\to e^{{1\over 2}a} A_s$ whereupon we find
that 
$$
{\cal E}= e^{-{\tau\over 2}}\tilde {\cal E}, \quad {\rm where }\quad 
\tilde {\cal E}= \left(\matrix {e_s&{1\over 2} e_s A_s\cr
0&e^{-1T}_s\cr}\right)
\eqno(11)$$
and $e^\tau= e^{-2a} det e_s  $. 
The Lagrangian of equation (10) then becomes 
$$
L= e^\tau (c_1 \partial_S \tilde M_{PQ} \partial_T (\tilde
M^{-1})^{PQ}(\tilde M^{-1})^{ST}+c_2 \partial_S (\tilde
M^{-1})^{PQ}\partial_P \tilde M_{TQ} (\tilde M^{-1})^{ST}
$$
$$
+c_3 \partial_S (\tilde M^{-1})^{SR}
\partial_P (\tilde M^{-1})^{PQ} \tilde M_{RQ})
-(20c_1+c_2 -c_3+c_4-2c_5)e^\tau \partial_S \tau \partial_T\tau \tilde
(M^{-1})^{ST} 
$$
$$
-2e^\tau (c_2-c_3- c_5) \partial_S\tau \partial_T (\tilde
M^{-1})^{ST}
\eqno(12)$$
where $\tilde M= \tilde {\cal E}\tilde {\cal E}^T$. 
We note that although ${\cal E}$ does not have
determinant one, 
$\tilde {\cal E}$ does have determinant one and is an element of
SO(10,10). 
\par
The field redefinition could also have been achieved by using a different
choice of generators in the group element. Indeed, we can  rewrite the
group element  of equation (3) in terms of the generators of SO(10,10),
given in appendix A,  and
$\tilde R$ to find 
$$
g_E= e^{h_s\cdot K} e^{{1\over 2} A_s\cdot R} e^{{1\over 3} a\tilde R} 
=e^{h_s\cdot \tilde K} e^{{1\over 2} A_s\cdot R} e^{{1\over 3}
\tilde a\tilde R} 
\eqno(13)$$
where $e_s=e^{h_s}$, as a matrix,  and $e^{-2\tilde a}=e^{\tau}$. Thus
string frame is associated with the group element when written in terms
of the SO(10,10) generators and $e^\tau$ is associated with the 
GL(1) factor. 
\par
So far we have constructed a theory which is
invariant   under a rigid symmetry namely 
$SO(10,10)) \otimes GL(1)$ but we would like to construct a  
theory that has local symmetries which replace the rigid symmetries of
$SO(10,10))$. This is a theory with 
diffeomorphisms and gauge symmetries. The way to do this has
been set out in the context of other theories. For the case of gravity
one writes down the non-linear realisation of $GL(D)\otimes_s P$ where
$P$ are the translations and then demands that the theory admit
diffeomorphism, or equivalently,  admit a simultaneous realisation with
the  conformal group [11]. This fixes the constants in the Lagrangian
which has just the rigid symmetries of the non-linear realisation  and
one finds Einstein's theory of gravity. A similar  procedure was
followed at low levels for the
$E_{11}\otimes l_1$ non-linear realisation  to find maximal supergravity
theories but in these theories  one also finds that the theory was gauge
invariant corresponding to the presence of the gauge fields [1,2]. A more
subtle strategy was carried out  in the context of  maximal 
supergravity in four dimensions and its
$E_7$ symmetry [12]. An
$E_{7}\otimes_s l_1$  non-linear realisation was constructed
where in this case the $l_1$ representations is the fifty six
dimensional representation of $E_7$. It was shown [12] that the
corresponding action admitted a diffeomorphism symmetry if one neglected
the  coordinate dependence on forty nine of  the fifty six coordinates,
leaving a seven dimensional dependence, and fixed  the constants in the
action to a particular set of values.   In fact  this diffeomorphism
invariance required a contribution from the four dimensional metric
of the eleven dimensional supergravity theory in which the $E_7$
theory was embedded.  Unlike the case of gravity, the $E_7$ theory 
can  not admit a conformal symmetry that would imply the rigid
symmetries  of the non-linear realisation become local. 
\par
We now follow this  same strategy here. We restrict the derivatives in
the generalised space-time to be only in the $x^a$ directions. Carrying
out this step let us evaluate  the Lagrangian of equations (10) and   (12)
by  substituting the expression for  $\tilde {\cal E}$ 
 of equation (11), we find that 
$$
L= c_1e^\tau (2\partial_\mu g_{\rho\kappa}\partial^\mu g^{\rho\kappa}
-{1\over 2}\partial_\mu A_{\rho}{}^{\kappa}\partial^\mu A^{\rho}{}_\kappa 
-{1\over 2}\partial_\mu A^2_{\rho\kappa}\partial^\mu g^{\rho\kappa})
$$
$$
+c_2e^\tau (g^{\rho\mu}\partial_\mu g^{\nu\kappa} \partial_\nu
g_{\rho\kappa}-{1\over 4}\partial^\rho A^\nu{}_\kappa \partial_\nu
A_\rho{}^\kappa+{1\over 4}\partial_\mu g^{\nu\kappa} \partial_\nu
A^\rho{}_\kappa A^\mu{}_\rho
+{1\over 4}\partial_\nu g^{\rho\kappa} \partial_\mu A^\nu{}_\kappa
A^\mu{}_\rho -{1\over 4}\partial^\rho g^{\nu\kappa}
\partial_\nu A^2_{\rho\kappa})
$$
$$
+c_3 e^\tau (\partial_\mu g^{\mu\rho} \partial_\nu g^{\nu\kappa}
g_{\rho\kappa}+{1\over 4} \partial_\mu A^\mu{}_\rho \partial_\nu
A^\nu{}_\kappa g^{\rho\kappa}  +{1\over 2} \partial_\mu A^\mu{}_\rho
\partial_\nu g^{\nu\kappa} A^\rho{}_\kappa 
-{1\over 4} \partial_\mu g^{\mu\rho} \partial_\nu
g^{\nu\kappa}A^2_{\rho\kappa})
$$
$$
-(20c_1+c_2 -c_3+c_4-2c_5)e^\tau \partial_\mu \tau \partial_\nu\tau
g^{\mu\nu} -2e^\tau (c_2-c_3- c_5) \partial_\mu\tau \partial_\nu 
g^{\mu\nu} 
\eqno(14)$$ 
where $\partial^\mu=g^{\mu\nu}\partial_\nu$ and $A^2_{\mu\nu}=
A_{\mu}{}^\rho A_\rho{}_\nu$. \par To fix the coefficients in the
Lagrangian it is simplest to carry out an infinitesimal diffeomorphism
and  gauge transformation on the bilinear terms and in particular the
variations   
$\delta h_{\mu\nu}= \partial_\mu \xi_\nu+ \partial_\nu\xi_\mu$,  where
$g_{\mu\nu}= e_\mu{}^a e_{\nu a}= \eta_{\mu\nu}+ 2h_{\mu\nu}$ at lowest
order,  and
$
\delta A_{\mu\nu} = \partial_\mu\Lambda_\nu-\partial_\nu\Lambda_\mu$ we
find the action is invariant provided 
$c_1=-{c_2\over 4},\ c_4=6c_2,\ c_5=0$. In fact the coefficient $c_2$
and $c_3$ occur in the combination $c_2-c_3$ and so we can set $c_3=0$.
We can choose the overall scale of the Lagrangian so that  
$$
c_1={1\over 4}, c_2=-1 , \ c_4 =-6
\eqno(15)$$
all others being zero. 
\par
We find that with these constants the Lagrangian is just that for the
NS-NS sector of the IIA supergravity which is given by 
$$
4\int d^{10} x \det e e^{-2a} \{ {1\over 2} R
-{1\over 3.8.4} F_{\mu_1\mu_2\mu_3}F^{\mu_1\mu_2\mu_3} +2\partial_\mu a
\partial^\mu a\}
\eqno(16)$$
In deriving this result we have used the  identity
$$
\int d^D x\sqrt {-det g} e^{-2a} R
$$
$$=\int d^D x\sqrt {-det g}e^{-2a}
\{-{1\over 2} \partial^\tau g^{\nu\lambda} \partial_\nu g_{\tau\lambda} 
+{1\over 4} \partial_\nu g_{\rho\kappa}  \partial^\nu g^{\rho\kappa}  
+{1\over 4} \partial_\nu (\ln\det g) \partial^\nu (\ln\det g)
$$
$$
-{1\over 2} \partial^\nu (\ln\det g)\partial^\mu g_{\mu\nu}
-2\partial_\nu a (2\Gamma^\nu -g^{\nu\tau} \partial^\rho g_{\tau\rho})
\}
$$
$$
= \int d^D x\sqrt {-det g}e^{-2a}
\{-{1\over 2} \partial^\tau g^{\nu\lambda} \partial_\nu g_{\tau\lambda} 
+{1\over 4} \partial_\nu g_{\rho\kappa}  \partial^\nu g^{\rho\kappa}  
+\partial_\mu\tau\partial^\mu\tau-\partial^\mu\tau
\partial^\lambda g_{\mu\lambda}-4\partial^\mu a\partial_\mu a\}
\eqno(17)$$ 
valid in $D$ dimensions and  
the definitions   
$\Gamma_{\mu\nu}^\lambda={1\over 2} g^{\lambda\tau}(\partial_\nu
g_{\tau\mu}+\partial_\mu g_{\tau\nu}-\partial_\tau g_{\mu\nu})$ and 
$R_{\mu\nu}{}^\rho{}_\lambda= \partial_\mu \Gamma_{\nu\lambda}^\rho
+\Gamma_{\mu\kappa}^\rho \Gamma _{\nu\lambda}^\kappa-(\mu\to \nu)$. 
\par 
It is interesting  to trace how the  familiar  density factor $\det e $
arises in the above Lagrangians. Each factor of $M$
carries with it a
$(\det e)^{-1}$ factor which it inherited from the product of two ${\cal
E}$'s each with factor $(\det e)^{-{1\over 2}}$, this  in turn had its
origin in  the last  terms, with factor of one half,  in the commutation
relations of
$K^a{}_b$ with
$P_a$ and $Q^a$ given in equation (2). This additional term arises as
$P_a$ is the highest weight state in the
$E_{11}$ representation. The two space-time derivatives
in the Lagrangian ensure that there is one more factor of $M^{-1}$ than
of $M$ in order to balance the indices and as a consequence we find  the
desired 
$\det e $ factor. It is interesting to  note that  this factor has its
origins in the $E_{11}$ algebra and the definition of the  $l_1$
representation. 
\par
 We now comment on the relation of this paper to the work
on the so called double field theory. This result was developed
in a number of substantial papers [17-20]    which had
their origins in earlier work on string field theory [21] and reference
[22]. The final  result was  a field theory with a metric, two form and
dilaton, i.e. the fields of the  massless NS-NS sector of closed
strings, but which depend on  the coordinates
$x^a$ and $y_a$ which transformed as a  vector of O(D,D). As
the author of this paper understands it,  doubled field theory was
found by a circuitous route beginning [17] by extracting  the
quadratic  and cubic terms,  together with their gauge transformations, 
from  gauge covariant   closed string field theory [23].  These terms
were then shown to be invariant under a set of O(10,10) transformations
introduced by ansatz. By assuming a set of requirements,  gauge
transformations were found to all orders in the fields  and written in an
O(10,10) covariant way [18]. By introducing covariant derivatives an all
orders Lagrangian was found that agreed with that of closed string field
theory at quadratic and cubic order [19] and shown to be invariant under
the gauge transformations if a constraint held. In the general case this
constraint was the same as suppressing all dependence on $y_a$. 
\par
Finally,  this  Lagrangian  was expressed in
terms of  an O(10,10) generalised metric $H$ (the
$\tilde M$ in this paper) and $e^{-2d}$ (the
$e^{-\tau}$ in this paper). 
In fact this Lagrangian was previously contained
in a seminar [24] of one of the authors (OH).  However, it was also
claimed in  the seminar that $H^{-1} \partial H$ was the
Cartan form of O(10,10). It was pointed out [25]  by  the author 
of this paper  that the final result was almost certainly the lowest
level non-linear realisation of $E_{11}\otimes _s l_1$, i.e. the result of
this paper, and that  the non-linear realisation would
contain   $\tilde M=gg^T$ where
$g\in O(10,10)$ with the transformation $g\to g_0gh$ being understood.
This relation  subsequently appeared in paper [20] but in the form 
$\tilde M=g^Tg$. Unfortunately this latter expression  is not inert
under the local $h$ part to the transformation $g\to g_0gh$ of
equation (5.1) of paper [20].  
\par
Although our  final result of equation (10) with the
constants of equation (15) agrees with that of [20] the derivation
presented in this paper is very different. We begin with an entirely
group theoretic construction, i.e the non-linear realisation of 
$E_{11}\otimes _s l_1$, as proposed in reference [3],  at lowest level.
It is important to realise that this is not the same as the
traditional sigma model associated with internal symmetries. 
The fields,   their rigid 
$O(D,D)\otimes GL(1)$ transformations and the most general  Lagrangian
are found by a very straightforward calculation. The constants in the
Lagrangian are fixed, following the ideas of [11,1,2,12], by demanding
that  the Lagrangian be diffeomorphsim invariant when the fields are
restricted to depend only on
$x^\mu$. Crucial to this construction is the GL(1) factor in the symmetry
which seems to play no  role in the papers on double field theory.
Such an additional factor is required from the non-linear
realisation perspective since the generators not in the local
subalgebra lead to fields in the final theory and we require a 
$D+1$ commuting subalgebra in order to find   the
diagonal components of the metric and the dilaton. As explained above it
is  also  essential to get the usual $\det e$  density
factor  in the Lagrangian.    
\par
The $E_{11}\otimes _s l_1$ non-linear realisation also provides a clear
method to include the Ramond-Ramond  fields and indeed higher
level  fields  and also the corresponding additional coordinates.
Furthermore it makes it clear that the symmetry underlying the doubled
field theory,  which had its origins in string field theory,   
is just that of the  non-linear realisation of
$E_{11}\otimes _s l_1$. 
\par 
It is instructive to  consider the eleven dimensional
theory formed from the $E_{11}\otimes _s l_1$ non-linear realisation with
local subalgebra $K(E_{11})$  at lowest level, i.e. level zero with
respect to the deletion of node eleven.  The resulting subalgebra is
GL(11) which is associated with eleven dimensional gravity. At level 
zero one has   the GL(11) algebra,  with generators
$K^a{}_b,\ a,b=1,\ldots , 11$ and the local subalgebra is just SO(10),
while in the
$l_1$ representation there are only  the usual space-time translations
$P_a$. Their commutation  relations,  as derived from $E_{11}$,  are given
by  
$$
[K^a{}_b,K^c{}_d]=\delta _b^c K^a{}_d - \delta _d^a K^c{}_b,  \ \ 
[K^c{}_b, P_a]=-\delta_a^c P_b+{1\over 2} \delta ^c_bP_a
\eqno(18)$$
We note the presence of the second 
 term in the last commutator whose origin was discussed in appendix A. We
can now follow exactly the path as above, we take the group element
$g=g_lg_E= e^{x\cdot P}e^{h\cdot K}$  and find from the Cartan form that 
${\cal E}_\mu {}^a = (dete )^{-{1\over 2}}e$ where
$e_\mu{}^a=(e^h)_\mu{}^a$. The most general invariant action is formed
from $\partial_\mu ={\partial\over \partial x^\mu}$ and $M={\cal E}{\cal
E}^T$ and has the above form of equation (10)  with the
indices
$M,N,\ldots$ now being  replaced by   $\mu,\nu=1,\ldots , 11$. We
find a diffeomorphism invariant result,  which is general relativity,  in
$D$ dimensions provided
$c_1={1\over 4},\ c_2=-{1\over 2},\ c_3=0=c_5$ and $ c_4=-{1\over D-2}$.
Of course $D=11$ for the case of interest to us.  However, we note that
the
$\det e$ factor  has the same origin as that just described above, namely
in the second term of the last commutator of equation (18). 
\par
It is amusing to consider the next level in eleven dimensions. At
level one we have the three form generator $R^{abc}$ in the adjoint
representation of $E_{11}$ and the generators $Z_{ab}$ in $l_1$
representation. Thus we have the field $A_{abc}$ which now depends on the
coordinates $x^a$ and $y_{ab}$. The group element is $g=g_lg_E$ with 
$g_l=e^{x\cdot P}e^{y\cdot Z}$ and $g_E= e^{h\cdot K}e^{A\cdot R}$. The
Cartan form takes the form  
$$
{\cal V}= g^{-1} d g= g_E^{-1}(dx^aP_a+ dy_{ab}Z^{ab}) g_E+ g_E^{-1}d g_E
$$
$$=(dx^a, dy_{ab}) {\cal E}  \left( \matrix {P_b\cr Z^{ab}}\right)
+ g_E^{-1}d g_E
\eqno(19) $$ 
where the generalised vierbein  ${\cal E}$ is given as a matrix by 
$$
{\cal E}= (det e)^{-{1\over 2}}
\left(\matrix {e_\mu{}^a&-{1\over 2} e_\mu{}^c A_{cb_1b_2}\cr
0&(e^{-1})_{[b_1}{}^{\mu_1} (e^{-1})_{b_2]}{}^{\mu_2}   \cr}\right)
\eqno(20)$$
One can then construct $M={\cal
E}{\cal E}^T$ which is invariant under local transformations. We hope to
report elsewhere on the dynamics of the  non-linear
realisation. The same calculation can easily be carried out for the IIA
theory at the next level. The additional fields and coordinates were
listed earlier in this paper. 
\par
In a previous paper [15] it was shown that quantising the SO(10,10)
symmetric string leads to a field theory with coordinates $x^a$,  or $y^a$
or a field theory with the coordinates $x^a$ and  $y^a$but which obey
non-trivial commutation relations, that is a  non-commutative field
theory. The different theories corresponding to the different choices of
representations of the fundamental commutators of the theory. It is
tempting to assume that all these different theories lead to the same
physical result, but it can not be excluded that they may differ in
subtle ways. It would be interesting to investigate the connection with
the results found in this paper. It is striking that some non-linear
realisations  admit, with a suitable choice of constants, local
symmetries. The deeper meaning of this result has yet to be understood. 

\medskip
{\bf Acknowledgment} 
\medskip
The author wishes to thank Christian Hillmann for extensive discussions 
that substantially helped the development of this work and  the 
Erwin Schr{\" o}dinger International Institute for
Mathematical Physics, Vienna for providing financial support and
hospitality during the author's stay in Vienna where much of this work was
carried out. He also thanks the Physics department of the Technical
University of Vienna for their hospitality and the STFC for support from
the rolling grant awarded to King's.


\medskip
{\bf Appendix A} 
\medskip
In this appendix  we calculate the algebra 
$E_{11}\otimes_s l_1$ at lowest level for the decomposition appropriate
to the IIA theory.  We
will begin with the  known algebra of the
$E_{11}$ algebra in terms of $A_{10}$, or SL(11), representations
[1,3]. This algebra is found by deleting node eleven in the Dynkin
Diagram of
$E_{11}$ ( see figure 1). For a decomposition of $E_{11}$
corresponding to a subalgebra associated with the deletion of a
particular node, the resulting generators 
can be classified in terms of increasing level. As for any Kac-Moody
algebra,  the  generators of  $E_{11}$ are constructed as multiple
commutators of the Chevalley generators and the
 level of a given positive (negative)  root generator with respect to a
particular node is just plus (minus) the number of times that the positive
(negative) root Chevalley generators associated with this node occurs in
this multiple commutator. For the decomposition to SL(11) we consider the
level associated with node eleven. At level zero we have the algebra 
GL(11) with the generators
$K^a{}_b,\ a,b =1,\ldots 11$ and at level one and minus one the rank
three generators $R^{abc}$ and 
$R_{abc}$. These level zero and one 
 generators contain  all the Chevalley generators of $E_{11}$ (the
positive root Chevalley generator associated with node eleven is
$R^{91011}$) and so their multiple commutators lead to all generators of
$E_{11}$. As a result the level of a 
generator that occurs in the SL(11) decomposition is just the number of
times the generator
$R^{abc}$ occurs  minus the number of times the generator $R_{abc}$
occurs. The generators at level two and minus two are
$R^{a_1\ldots a_6}$
 and $R_{a_1\ldots a_6}$ respectively, while those at levels three and
minus three are $R^{a_1\ldots a_8,b}$
 and $R_{a_1\ldots a_8,b}$ respectively
\par
The $E_{11}$ algebra at   levels zero and up  three is
given by [1,3]
$$
[K^a{}_b,K^c{}_d]=\delta _b^c K^a{}_d - \delta _d^a K^c{}_b, Ê
\eqno(A.1)$$
$$Ê [K^a{}_b, R^{c_1\ldots c_6}]=Ê
\delta _b^{c_1}R^{ac_2\ldots c_6}+\dots, \ Ê
Ê[K^a{}_b, R^{c_1\ldots c_3}]= \delta _b^{c_1}R^{a c_2 c_3}+\dots,
\eqno(A.2)$$
$$ [ K^a{}_b,Ê R^{c_1\ldots c_8, d} ]=Ê
(\delta ^{c_1}_b R^{a c_2\ldots c_8, d} +\cdots) + \delta _b^d
R^{c_1\ldots c_8, a} .
\eqno(A.3)$$
and 
$$[ R^{c_1\ldots c_3}, R^{c_4\ldots c_6}]= 2 R^{c_1\ldots c_6},\quad 
[R^{a_1\ldots a_6}, R^{b_1\ldots b_3}]
= 3Ê R^{a_1\ldots a_6 [b_1 b_2,b_3]},Ê
\eqno(A.4)$$
where $+\ldots $ means the appropriate anti-symmetrisation.Ê

The $E_{11}$ level zero  and negative level generators up to level minus
three obey the relationsÊ
$$
[K^a{}_b, R_{c_1\ldots c_3}]= -\delta ^a_{c_1}R_{b c_2
c_3}-\dots,\ [K^a{}_b, R_{c_1\ldots c_6}]=Ê -\delta ^a_{c_1}R_{bc_2\ldots
c_6}-\dots,
\eqno(A.5)$$
$$ [ K^a{}_b,Ê R_{c_1\ldots c_8, d} ]=Ê
-(\delta ^a_{c_1} R_{b c_2\ldots c_8, d} +\cdots) - \delta ^a_d
R_{c_1\ldots c_8, b} .
\eqno(A.6)$$
$$[ R_{c_1\ldots c_3}, R_{c_4\ldots c_6}]= 2 R_{c_1\ldots c_6},\quadÊ
[R_{a_1\ldots a_6}, R_{b_1\ldots b_3}]
= 3Ê R_{a_1\ldots a_6 [b_1 b_2,b_3]},Ê
\eqno(A.7)$$
Finally, the commutation relations between the positive and negative
generators Ê are given byÊ

$$[ R^{a_1\ldots a_3}, R_{b_1\ldots b_3}]= 18 \delta^{[a_1a_2}_{[b_1b_2}
K^{a_3]}{}_{b_3]}-2\delta^{a_1a_2 a_3}_{b_1b_2 b_3} D,\ Ê
[ R_{b_1\ldots b_3}, R^{a_1\ldots a_6}]= {5!\over 2}
\delta^{[a_1a_2a_3}_{b_1b_2b_3}R^{a_4a_5a_6]}
$$
$$
[ R^{a_1\ldots a_6}, R_{b_1\ldots b_6}]= -5!.3.3
\delta^{[a_1\ldots a_5}_{[b_1\ldots b_5}
K^{a_6]}{}_{b_6]}+5!\delta^{a_1\ldotsÊ a_6}_{b_1\ldotsÊ b_6} D ,\quadÊ
$$
$$
[ R_{a_1\ldots a_3}, R^{b_1\ldots b_8,c}]= 8.7.2
( \delta_{[a_1a_2 a_3}^{[b_1b_2b_3} R^{b_4\ldots b_8] c}-
Ê\delta_{[a_1a_2 a_3}^{[b_1b_2 |c|} R^{b_3\ldots b_8]} )
$$
$$
[ R_{a_1\ldots a_6}, R^{b_1\ldots b_8,c}]= {7! .2\over 3}
( \delta_{[a_1\ldotsÊ a_6}^{[b_1\dots b_6} R^{b_7 b_8] c}-
Ê\delta_{[a_1\ldotsÊ a_6}^{c[b_1\ldots b_5 } R^{b_6b_7 b_8]})
\eqno(A.8)$$
where $D=\sum_b K^b{}_b$, $\delta^{a_1a_2}_{b_1b_2}=
{1\over
2}(\delta^{a_1}_{b_1}\delta^{a_2}_{b_2}-
\delta^{a_2}_{b_1}\delta^{a_1}_{b_2})=
\delta^{[a_1}_{b_1}\delta^{a_2]}_{b_2}$ with similar formulae whenÊ
more indices are involved.ÊWe have taken the liberty of listing the
algebra to a somewhat higher than required in this paper. In fact
these equations correct the coefficients of one of the equations contained
in reference [3]. 
\par
The IIA theory arises  when we consider the deletion of node ten in the
$E_{11}$ Dynkin diagram to leave a SO(10,10) algebra leading to a
decomposition of $E_{11}$ into representations of this algebra. However,
it is  illuminating to then delete node eleven and analyse the
representations of SO(10,10) in terms of SL(10). Thus in this case we have
two levels which we may write as $(l_1,l_2)$ and  are associated with
nodes ten and eleven respectively. At level $l_1=0$  we find 
generators with $l_2=0$ which are the  generators  of GL(10), denoted by
$K^a{}_b,
\ a,b=1,2\ldots ,10 $,  and the generator of $\tilde R$  which is some
combination of $\sum_{a=1}^{11} K^a{}_a$ and $K^{11}{}_{11}$ as well as
the generators $ R^{ab11}\equiv  R^{ab}$ and $ R_{ab}\equiv R_{ab 11}$ at
levels $l_2=1$ and $l_2=-1$ respectively. Substituting into the above
algebra of equations (A.1-A.8) one finds that these generators obey the
algebra 
$$
[K^a{}_b,K^c{}_d]=\delta _b^c K^a{}_d - \delta _d^a K^c{}_b,  \ \ 
 [K^a{}_b, R^{c d}]=\delta_b^c R^{ad}-\delta_b^d R^{ac},
[K^a{}_b,  R_{c d}]=-\delta_c^a \tilde R_{bd}+\delta_d^a \tilde
R_{bc},
$$
$$
 [R^{ab},  R_{c d}]=4\delta_{[c}^{[a}
K^{b]}{}_{d]}+{2\over 3} \tilde  R,\ \ \ [R^{ab}, R^{c d}]=0
=[\tilde R_{ab}, \tilde R_{c d}] , \ [\tilde R, R^{ab~}]=0,\  \ [\tilde
R, R_{ab~}]=0
\eqno(A.9)$$
where $\tilde R= - \sum_{a=1}^{11} K^a{}_a +3 K^{11}{}_{11}$. 
\par
At first sight this is not obviously the algebra of 
$SO(10,10) \otimes GL(1)$, 
however, if we redefine the GL(10) generators by 
$$
\tilde K^a{}_b= K^a{}_b +{1\over 6}\delta_a^b \tilde R  \ ,
\eqno(A.10)$$
leaving all other generators the same one,  finds the above algebra
becomes 
$$
[\tilde K^a{}_b,\tilde K^c{}_d]=\delta _b^c \tilde K^a{}_d - \delta _d^a
\tilde K^c{}_b,  \ \ 
 [\tilde K^a{}_b, R^{c d}]=\delta_b^c R^{ad}-\delta_b^d R^{ac},
[\tilde K^a{}_b,  R_{c d}]=-\delta_c^a \tilde R_{bd}+\delta_d^a \tilde
R_{bc},
$$
$$
 [R^{ab},  R_{c d}]=4\delta_{[c}^{[a}
\tilde K^{b]}{}_{d]},\ \ \ [R^{ab}, R^{c d}]=0=[\tilde R_{ab}, \tilde R_{c d}]
\eqno(A.11)$$
which we recognise as the algebra of SO(10,10) and an 
additional generator $\tilde R$ that commutes with all the SO(10,10)
generators. 
\par
We also need the fundamental representation of $E_{11}$  associated with
node one. By definition this is the representation with highest weight
$\Lambda_1$ which obeys 
$(\Lambda_1, \alpha_{  a})=\delta _{a,1}, \  a=1,2\ldots ,11$ 
where $\alpha_{ a}$ are the simple roots of $E_{11}$. In the
decomposition to Sl(11), corresponding to the deletion of node eleven, 
one finds that the $l_1$ representation  contains  the objects $P_a,\ 
a=1,\ldots , 11$,
$Z^{ab}$ and
$Z^{a_1\ldots a_5}$ corresponding to levels zero, one and two
respectively. We have taken the first object, i.e. $P_a$,  to have level
zero by choice. Taking these to be generators in a semi-direct product
group denoted
$E_{11}\otimes _s l_1$ their commutation relation with the level one
generators of $E_{11}$ are given by 
$$
[R^{a_1a_2a_3}, P_b]= 3 \delta^{[a_1}_b Z^{a_2a_3]}, \ Ê
[R^{a_1a_2a_3}, Z^{b_1b_2} ]= Z^{a_1a_2a_3 b_1b_2},\ Ê
$$
$$[R^{a_1a_2a_3}, Z^{b_1\ldots b_5} ]=Z^{b_1\ldots b_5[a_1a_2,a_3]}+
Z^{b_1\ldots b_5 a_1a_2 a_3}
\eqno(A.12)$$ 
These equations define the normalisation 
of the generators of the $l_1$ representation.ÊThe commutator of
$K^a{}_b$ with $P_c$ can only take the form 
$[K^a{}_b, P_c]= -\delta _c^a P_b +e\delta _b^a P_c$ where $e$ is a
constant. This commutator was found [3] to have $e={1\over 2}$ as a result
of the fact that $l_1$ is an $E_{11}$ representation. We summarise the
argument here as this relation plays a crucial role in this paper. Since  
$P_1$ corresponds to the highest weight state  in the representation the
action of the Chevalley generator $H_{11}$ is given by 
$[H_{11}, P_1]= (\alpha_{11},\Lambda_1 ) P_1 $  where
$H_{11}= {2\over 3}(K^9{}_9+K^{10}{}_{10}+K^{11}{}_{11}) -{1\over
3}(K^1{}_1+\dots +K^{8}{}_{8})$ is the Chevalley generator in the Cartan
subalgebra associated with node eleven. Since $(\Lambda_1 ,
\alpha_{11})=0$ we must have $[H_{11}, P_1]=0$ and then we find $e={1\over
2}$ as claimed.  Using  the Jacobi identity and the fact thatÊ $e={1\over
2}$ we conclude thatÊ
$$
Ê[K^a{}_b, P_c]= -\delta _c^a P_b +{1\over
2}\delta _b^a P_c,\ Ê [K^a{}_b, Z^{c_1c_2} ]= 2\delta_b^{[c_1} Z^{|a|c_2]}
+{1\over 2}\delta _b^a Z^{c_1c_2},
$$
$$
[K^a{}_b, Z^{c_1\ldots c_5} ]= 5\delta_b^{[c_1} Z^{|a|c_2\ldots c_5]}
+{1\over 2}\delta _b^a Z^{c_1\ldots c_5}
\eqno(A.13)$$
Using the Jacobi identities, equations (A.13)
and equations (A.1-A.8)Ê we find thatÊ
$$
[R^{a_1\dots a_6}, P_b]= -3 \delta^{[a_1}_b Z^{\ldots a_6]}, \ 
[R^{a_1\dots a_6}, Z^{b_1b_2} ]= Z^{b_1b_2[a_1\ldots a_5,a_6]},\ Ê
\eqno(A.14)$$
and theÊ commutators with the negative root generatorsÊare given by 
$$
[R_{a_1a_2a_3}, P_b ]= 0,\ Ê
[R_{a_1a_2a_3}, Z^{b_1b_2} ]= 6\delta^{b_1b_2 }_{[a_1a_2} P_{a_3 ]},\ Ê
[R_{a_1a_2a_3}, Z^{b_1\ldots b_5} ]= {5!\over 2} \delta^{[ b_1b_2b_3
}_{a_1a_2a_3} Z^{b_4b_5]}
\eqno(A.15)$$
\par
As seen from the IIA perspective we first delete node  ten and then node
eleven; the level $l_1=0$ generators of the $l_1$ multiplet are $P_a, 
\  a=1,\ldots , 10$ and $Q^a\equiv -2Z^{a11}$ which have levels $l_2=0$
and 
$l_2=1$ respectively. We find from the above algebra  that they obey the
commutation relations 
$$
[K^c{}_b, P_a]=-\delta_a^c P_b+{1\over 2} \delta ^c_bP_a,\ \ 
 [  R^{ab}, P_c]=-{1\over 2}
(\delta^a_c Q^b-\delta^b_c Q^a)=- \delta^{[a}_c Q^{b]}, \ \  [\tilde 
R_{ab}, P_c]=0,
$$
$$
[K^a{}_b, Q^c]=\delta_b^c Q^a+{1\over 2} \delta ^c_b Q^a, \ \  [ \tilde
R_{ab},  Q^c]=2 (\delta_a^c P_b-\delta_b^c P_a)= 4 \delta_{[a}^c P_{b]},\
\  [R^{ab},Q^c]=0,
$$
$$
[\tilde R, P_a]= -3P_a,\ [\tilde R, Q^a]= -3Q^a
\eqno(A.16)$$
It will be useful to also give the commutators in terms of the
generator
$R={1\over 12} (-\sum_{a=1}^{11} K^a{}_a +9K^{11}{}_{11})$, used in the
non-linear realisation,  which leads to 
$$
[R, P_a]= 0,\ [R, Q^a]= {1\over 2}Q^a , \ [R, R^{ab}]={1\over 2}R^{ab} , 
\ [R, R_{ab}]=-{1\over 2}R_{ab} 
\eqno(A.17)$$
\par
It is instructive to give these relations when we take the generators  of
equation (A.10) that are those of the SO(10,10) algebra. Then one finds
that 
$$
[\tilde K^c{}_b, P_a]=-\delta_a^c P_b,\ \ 
 [  R^{ab}, P_c]=-{1\over 2}
(\delta^a_c Q^b-\delta^b_c Q^a)=- \delta^{[a}_c Q^{b]},\ [\tilde
R_{ab},P_c]=0,\ 
$$
$$
[\tilde K^a{}_b, Q^c]=\delta_b^c Q^a ,
 \  [ \tilde R_{ab},  Q^c]=2
(\delta_a^c P_b-\delta_b^c P_a)= 4 \delta_{[a}^c P_{b]},\
\  [R^{ab},Q^c]=0,
\eqno(A.18)$$
which are those corresponding to the usual vector representation of
SO(10,10). The relations with the additional GL(1) factor are given in
equation (A.16).
\medskip
$$
\matrix{
& & & & &&& &\bullet &11&&&
\cr & & & &&& & &| & && &
\cr
\bullet&-&\bullet&-&\ldots &- &\bullet&-&\bullet&-&\bullet&-&\bullet
\cr
1& &2& & & &7& &8& & 9&
&10\cr}
$$
\par
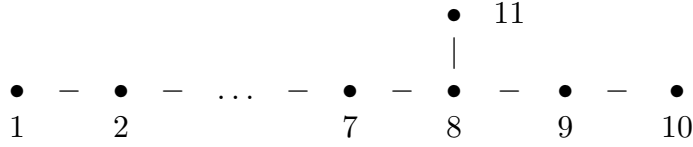
\centerline {Fig 1. The $E_{11}$ Dynkin diagram}

\medskip


\medskip
{\bf References}
\medskip
 \item{[1]} P. West, {\sl $E_{11}$ and M Theory}, Class. Quant.
Grav. {\bf 18 } (2001) 4443, {\tt hep-th/0104081}
\item{[2]} P.~C. West, {\sl Hidden superconformal symmetry in {M}
    theory },  JHEP {\bf 08} (2000) 007, {\tt hep-th/0005270}
\item{[3]} P. West, {\sl $E_{11}$, SL(32) and Central Charges},
Phys. Lett. {\bf B 575} (2003) 333-342, {\tt hep-th/0307098}
\item{[4]} P. West, {\sl $E_{11}$ origin of brane charges and U-duality
  multiplets}, {\bf  JHEP} 0408 (2004) 052, hep-th/0406150.
\item{[5]} A. Kleinschmidt and  P. West, {\sl
Representations of
${\cal G}^{+++}$ and the role of space-time}, JHEP {\bf 0402} (2004) 033,
hep-th/0312247. 
\item{[6]} P. West, Brane dynamics, central charges and $E_{11}$, JHEP
0503 (2005) 077, hep-th/0412336. 
\item{[7]} P. Cook and P. West, Charge multiplets and masses for E(11); 
JHEP 11 (2008) 091,  arXiv:0805.4451.
 \item{[8]} F. Riccioni and P. West, E(11)-extended space-time and gauged
supergravities, JHEP0802:039,2008;  hep-th/0712.1795
\item{[9]} N. Hitchin, Generalized Calabi-Yau manifolds, 
 Q. J. Math.  {\bf 54}  (2003), no. 3, 281,
math.DG/0209099;  Brackets, form and
invariant functionals, math.DG/0508618.
\item{[10]} M. Gualtieri, Generalized complex geometry, PhD Thesis
(2004), math.DG/0401221v1. 
\item {[11]} Borisov and V. Ogievetsky, Theory of 
dynamical affine and conformal symmetries as gravity theory  of
  the gravitational field,  Theor.\ Math.\ Phys.\  {\bf 21} (1975) 1179. 
\item {[12]]} C. Hillmann, Generalized E(7(7)) coset dynamics and D=11
supergravity, JHEP {\bf 0903}, 135 (2009), hep-th/0901.1581.
\item{[13]} M. Duff, Duality Rotations In String Theory,
  Nucl.\ Phys.\  B {\bf 335} (1990) 610; M. Duff and J. Lu,
 Duality rotations in
membrane theory,  Nucl. Phys. {\bf B347} (1990) 394. 
\item{[14]} A. Tseytlin, Duality Symmetric Formulation Of String World
Sheet Dynamics, Phys.Lett. {\bf B242} (1990) 163, Duality Symmetric
Closed String Theory And Interacting Chiral Scalars, Nucl.\ Phys.\ B {\bf
350}, 395 (1991). 
\item {[15]} P. West, Generalised space-time and duality,
hep-th/1006.0893.  
\item {[16]} I. Schnakenberg and P. West,  Kac-Moody Symmetries of IIB
Supergravity,  Phys.Lett. B517 (2001) 421-428, hep-th/0107181. 
\item{[17]} C. Hull and B. Zwiebach, Double Field Theory,  JHEP {\bf 0909}
(2009) 099, hep-th/0904.4664.
\item{[18]}  C. Hull and B. Zwiebach, The gauge algebra of double field
theory and Courant brackets,
  JHEP {\bf 0909} (2009) 090, hep-th0908.1792.
\item{[19]} O. Hohm, C. Hull and B. Zwiebach, Background independent
action for double field theory, hep-th/1003.5027.
\item{[20]} O. Hohm, C. Hull and B. Zwiebach, Generalised metric
formulation of double field theory,  hep-th/1006.4823. 
\item{[21]} T. Kugo and B. Zwiebach,Target space duality as a symmetry
of string field theory,
  Prog.\ Theor.\ Phys.\  {\bf 87}, 801 (1992) hep-th/9201040.
\item{[22]}  W. Siegel, Superspace duality in low-energy superstrings,
  Phys.\ Rev.\  D {\bf 48}, 2826 (1993), hep-th/9305073;
  Two vierbein formalism for string inspired axionic gravity,
  Phys.\ Rev.\  D {\bf 47}, 5453 (1993) hep-th/9302036.
\item {[23]} T. Kugo and B. Zweibach, Prog. Theor. Phys. {bf 87}, (1992)
801.  
\item{[24]} Seminar of Olaf Holm given at the  Solvay workshop, 
"Symmetries and dualities in gravitational  theories"  in Brussels, May
21,  2010.
\item{[25]} Discussion session after the seminar of Olaf Holm specified
in  the  previous reference.


\end